 \documentclass[smallabstract,smallcaptions]{dccpaper}

\usepackage{epsfig}
\usepackage{amsmath}
\usepackage{amssymb}
\usepackage{color}
\usepackage{url}

\newlength{\figurewidth}
\newlength{\smallfigurewidth}

\begin{document}

\title
{\large
\textbf{A Preprocessing Framework for Video Machine Vision under Compression}
}

\author{%
Fei Zhao$^{\ast}$\thanks{$^{1}$Intern in Bytedance. $^{2}$Corresponding author.}\textsuperscript{1}, Mengxi Guo$^{\dag}$, Shijie Zhao$^{\dag}$\textsuperscript{2}, Junlin Li$^{\dag}$, Li Zhang$^{\dag}$ and Xiaodong Xie$^{\ast}$\\[0.5em]
{\small\begin{minipage}{\linewidth}\begin{center}
\begin{tabular}{ccc}
$^{\ast}$School of Computer Science, & \hspace*{0.5in} & $^{\dag}$Bytedance \\
$^{\ast}$Peking University && No. 48 Zhichun Road \\
No.5 Yiheyuan Road && Beijing, Beijing, 100098, China\\
Beijing, Beijing, 100871, China && \url{guomengxi.qoelab
@bytedance.com}\\
\url{feizhao@stu.pku.edu.cn} && \url{zhaoshijie.0526@bytedance.com}\\
\url{donxie@pku.edu.cn} && \url{lijunlin.li@bytedance.com}\\
&& \url{lizhang.idm@bytedance.com}\\
\end{tabular}
\end{center}\end{minipage}}
}

\maketitle
\thispagestyle{empty}

\begin{abstract}
There has been a growing trend in compressing and transmitting videos from terminals for machine vision tasks.
Nevertheless, most video coding optimization method focus on minimizing distortion according to human perceptual metrics, overlooking the heightened demands posed by machine vision systems.
In this paper, we propose a video preprocessing framework tailored for machine vision tasks to address this challenge.
The proposed method incorporates a neural preprocessor which retaining crucial information for subsequent tasks, resulting in the boosting of rate-accuracy performance.
We further introduce a differentiable virtual codec to provide constraints on rate and distortion during the training stage.
We directly apply widely used standard codecs for testing. Therefore, our solution can be easily applied to real-world scenarios.
We conducted extensive experiments evaluating our compression method on two typical downstream tasks with various backbone networks. The experimental results indicate that our approach can save over 15\% of bitrate compared to using only the standard codec anchor version.


\end{abstract}

\Section{Introduction}


With the success of deep learning, machine vision methods have made significant strides in recent years. Simultaneously, there has been a surge in video capture at edge devices, with subsequent transmission for machine vision tasks. The challenge for video compression optimization lies in reducing transmission bandwidth while upholding the accuracy for downstream machine vision tasks.


Despite the presence of several traditional video compression standards such as H.264/AVC\cite{h264} and H.265/HEVC\cite{HEVC}, these standards prioritize minimizing compression distortion for the human eyes rather than catering to the requirements of machine vision tasks. Furthermore, most standard codecs are not differentiable, making it difficult to do the end-to-end optimization with learning-based modules.  
Besides, the compression operation can result in a significant degradation of machine vision task performance.
Through experimental testing, we have observed a significant decline in machine vision performance, particularly under low-bitrate compression conditions. 

\begin{figure}[t]
\begin{center}
\includegraphics[width=15cm]{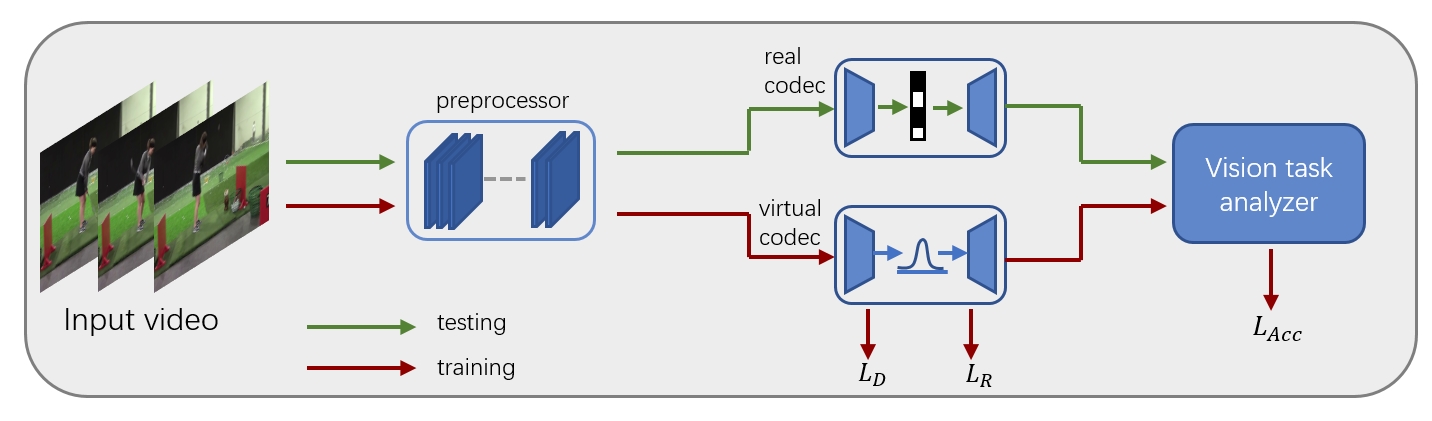} \\
\end{center}
\caption{\label{fig:pipeline}%
The overall pipeline of the proposed framework. During the training phase, we employ the virtual codec to generate the necessary distortion loss and rate loss for supervision, while accuracy loss is generated by the vision task. In the testing phase, we encode the videos processed by the preprocessor using a real standard codec to obtain the bitrate. We also collect the corresponding accuracy metrics from the downstream vision task to assess the overall performance.}
\end{figure}

In response to those limitations, we introduce a preprocessing framework tailored for machine vision in the context of compression, as illustrated in Figure \ref{fig:pipeline}. We propose a learning-based preprocessor that aims to enhance the performance of downstream vision tasks under compression without compromising perceptual quality for human viewing.
To facilitate end-to-end optimization, we further design a differentiable virtual video codec which emulates the distortion and rate for training supervision, and enabling gradient propagation to the learning-based preprocessor. 
During training, we utilize the virtual codec to generate rate-distortion loss, and the vision task analyzer contributes the accuracy loss. These three components of loss are combined as the optimization function during preprocessor training. In testing, we employ the real codec to evaluate the machine vision performance of the preprocessed and compressed videos, assessing them based on bitrate and accuracy.
We conduct extensive experiments on two prominent machine vision tasks, video action recognition and video object tracking, employing diverse downstream backbone networks. Our experiments reveal that, in comparison to traditional codecs like H.264/AVC and H.265/HEVC, our proposed approach achieves a bitrate savings of over 15\%.

Our work offers the following key contributions:

\begin{quote}
(1)We introduce a deep neural preprocessor designed to process input video and maintain a high level of downstream machine vision accuracy even when a substantial degree of compression is applied.
\end{quote}

\begin{quote}
(2)We propose a differentiable virtual video codec that enables the preprocessor to perceive the impact of compression during training.
\end{quote}

\begin{quote}
(3)Our experimental results showcase that our approach surpasses the anchor in terms of rate-accuracy performance. Furthermore, the robustness and universality of our method are noteworthy.
\end{quote}


\Section{Related Works}


\SubSection{Video Compression}
Video sequences typically contain a significant amount of information that can be considered redundant. Three main types of redundancies are commonly identified in video sequences: spatial redundancy, temporal redundancy, and statistical redundancy. These redundancies can be exploited to reduce the quantity of data required to represent the video sequence, improving compression efficiency and reducing storage requirements. 
Modern coding standards adopt the hybrid encoding scheme that builds upon these redundancy reduction techniques, which is the basic framework of standard codecs like H.264/AVC and H.265/HEVC.

\SubSection{Machine Vision with Compression}

Machine vision tasks differ from human vision tasks in terms of their objectives and evaluation metrics. Encoding video content for machine consumption presents an intriguing and challenging problem. 
The emergence of AI-driven video intelligent solutions play a pivotal role in addressing the most pressing challenges.
For instance, Torfason et al.\cite{torfason2018towards} introduced a deep neural compression framework that can be extended to jointly train networks for classification and segmentation tasks, facilitating image understanding.
Zhu et al.\cite{zhu2022perceptual} introduced a semantic-guided texture synthesis framework for video coding, with the objective of enhancing the compression efficiency of texture regions in videos.
The majority of existing approaches depend on learning-based compression methods to enable end-to-end optimization, which may not be feasible in real-world applications.

\SubSection{Preprocessing for video coding}
Previous studies have demonstrated that removing noise or insignificant details from video data can enhance the perceptual quality of video coding.
For instance, a preprocessing method which adaptively adjust filtering strength with the content of input video was proposed in \cite{xiang2016adaptive}.
Talebi et al.\cite{talebi2021better} applied pre-processing to images using a CNN before subjecting them to the JPEG encoding process to mitigate compression artifacts. Recently, Chadha et al.\cite{dpp} proposed a convolutional network-based perceptual preprocessing model to improve video coding performance.
However, existing works are predominantly designed for the human visual system, with limited attention given to downstream machine vision tasks.

\Section{Proposed Method}
\SubSection{Overview}
The pipeline of our proposed approach is depicted in Figure \ref{fig:pipeline}. During the training phase, we employ the preprocessor to manipulate the input video. The processed video then passes through our virtual codec to generate corresponding rate and distortion, serving as a part of the training supervision.
The distortion is determined by the difference between the reconstructed video from the virtual codec and the source input video, while the rate is estimated from the distribution of the reconstructed residuals. Details about the design of the virtual codec will be presented in the subsequent section.
The reconstructed video outputted by the virtual codec serves as input to the machine vision analyzer, from which accuracy metrics are obtained and included in the training supervision. 
The optimization function during training is presented as follows,

\begin{equation}
    L=\alpha(L_D + \lambda \times L_R) + L_{Acc} ,
\end{equation}
\noindent
where $L_{D}$ represents distortion loss, $L_R$ signifies rate loss, and $L_{Acc}$ denotes the accuracy loss pertaining to the respective downstream machine vision tasks. Note that the vision task analyzers are fixed during training, so we only update the parameters in the preprocessor for each training iteration. $\lambda$ is the weight coefficient balancing the trade-off between rate and distortion, while $\alpha$ signifies the weight coefficient between the compression-related rate-distortion loss and the accuracy of machine vision tasks. For each machine vision network, we individually trained a corresponding preprocessor to conduct our testing.

During testing, we apply the trained preprocessor, to process the source video. Subsequently, we use a real standard codec to compress the processed video, yielding corresponding distortion and rate values. We then utilize the compressed video for downstream machine vision tasks and assess its performance.

\SubSection{Preprocessing Network}

\begin{figure}[t]
\begin{center}
\includegraphics[width=5.5in]{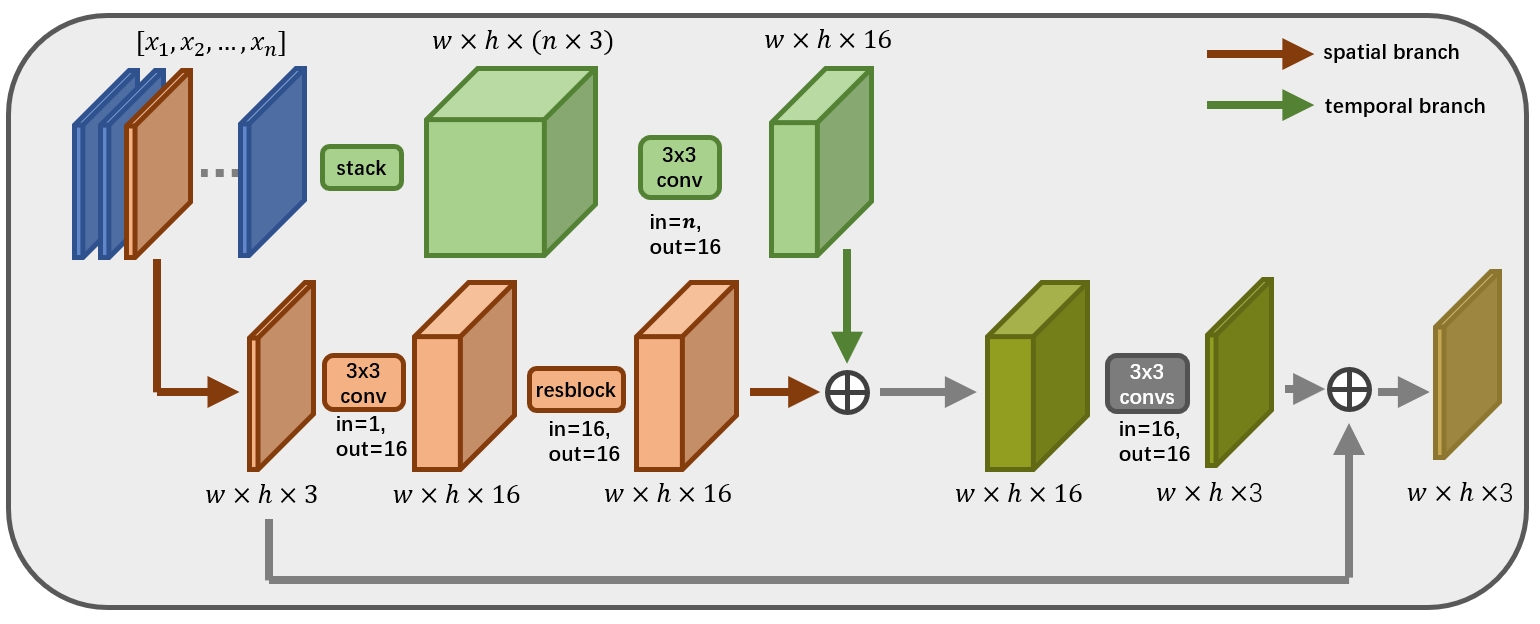} \\
\end{center}
\caption{\label{fig:proc}%
The structure of the proposed preprocessor can be divided into two primary branches, each responsible for extracting temporal and spatial features from the input video and subsequently merging them.}
\end{figure}

The objective of the preprocessing model is twofold: firstly, to perceive the impact of coding distortion throughout the entire pipeline, and secondly, to discern the characteristics of features required by downstream machine vision tasks under such conditions.
When confronted with downstream video machine vision tasks involving multi-frame inputs, our preprocessor must effectively harness inter-frame temporal information for video processing. Additionally, as the encoder eliminates temporal and spatial redundancies in the video, the preprocessor must be sufficiently sensitive to both the inter-frame temporal features and intra-frame spatial features of the video.

As depicted in Figure \ref{fig:proc}, we employed a convolutional neural network with multiple branch residual connections to constitute the preprocessing model.
Specifically, in the temporal branch, we conduct convolutions in the inter-frame dimension, yielding feature maps imbued with temporal information. In the spatial branch, we exclusively apply convolutions within individual frames, resulting in feature maps primarily composed of spatial information. Subsequently, we merge these two sets of feature maps using conditional attention and incorporate them into the original image using a residual connection, yielding the final output.

\SubSection{Virtual Video Codec}
Due to the non-differentiable nature of traditional video codecs, their integration into solutions based on neural network training for joint optimization is precluded. To address this limitation, we employ a virtual codec structure to stand in for the actual codec in the training process. The intention is to use this structure to establish constraints on distortion and rate, thereby constituting a part of the entire pipeline optimization function.

\begin{figure}[t]
\begin{center}
\includegraphics[width=5.5in]{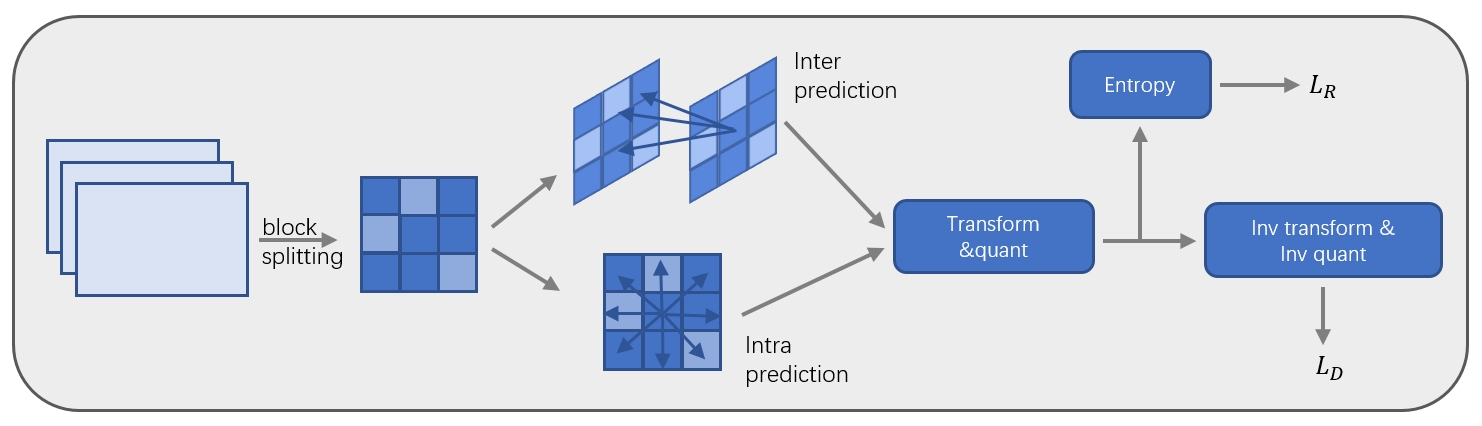} \\
\end{center}
\caption{\label{fig:vc}%
The proposed virtual codec emulates the fundamental logic used in video encoding and implements it through tensor operations. During the training phase, the virtual codec is responsible for providing rate loss and distortion loss.}
\end{figure}

The virtual video codec primarily generates distortion and rate by producing a residual map akin to that found in video codecs. We modeled the virtual codec as illustrated in Figure \ref{fig:vc}. Employing intra-frame and inter-frame (with the preceding frame as the reference frame) searches, we obtain the predicted residue for the current frame. Subsequently, transform, quantization, inverse quantization, and inverse transform are applied to get the reconstructed residual. Here, we employ quantaization factor $f_q$, an equivalent parameter representing quantization parameter (QP), to regulate the quantization strength. This reconstructed residual enables the restoration of the reconstructed frame, thus obtaining the corresponding distortion.
For rate computation, we utilize the factorized prior entropy method proposed by Balle et al.\cite{balle2016end}

\Section{Experiments}
\SubSection{Implementation}

Regarding the action recognition task, our framework is both trained and tested on the Kinetics400\cite{kay2017kinetics} dataset. We evaluate the performance of the proposed framework under various compression settings and report the top-1 to top-5 accuracy results. In our experiments, we employ six widely recognized action recognition networks, Slow-only\cite{feichtenhofer2019slowfast}, SlowFast\cite{feichtenhofer2019slowfast}, C2D\cite{c2d}, I3D\cite{i3d}, Swin\cite{swin} and TPN\cite{tpn} for the evaluation process.
In the context of video object tracking, we employ the GOT-10K\cite{got10k} dataset. Our experiments yield reported results in terms of Area under the ROC Curve (AUC) rates. To illustrate the efficacy of our approach, we utilize the backbones KYS\cite{kys}, DiMP\cite{dimp}, ATOM\cite{atom} and PrDiMP\cite{prdimp} for performance evaluation.

In our experiments, we utilize bits per pixel (bpp) as a measure to gauge the coding cost incurred during the compression process. For testing purposes, we utilized the widely used popular open-source implementations of H.264 and H.265, namely x264\cite{x264web} and x265\cite{x265web}, as standard codecs. The coding preset for testing is set to 'medium' and QP=[30,35,40,45,50]. The distortion loss in the loss function (1) is calculated using Mean Squared Error (MSE), while the rate loss is obtained by computing the estimated bpp. The weight parameter in loss function $\alpha$ and $\lambda$ are set to 10 and 0.001.

Our entire framework is developed using PyTorch\cite{imambi2021pytorch} with CUDA support and trained on A100 GPUs. The quantization factor $f_q$ of the virtual codec is set to randomly select values between 30 and 50. The weights of the downstream networks remain constant throughout the entire training process. Given that we solely need to train the preprocessing network within the complete framework, our training workflow requires a single end-to-end training iteration. We employ the Adam optimization method with a initial learning rate set to 1e-4.

\SubSection{Experimental Results}

\begin{table}[tp]
\begin{center}
\caption{\label{}%
BD-Rate(\%) performance for action recognition}
{
\begin{tabular}{|c|c|c|c|c|c|c|c|}
\cline{1-8}
\multicolumn{2}{|c|}{~}&
\multicolumn{6}{c|}{backbones}\\
\cline{1-8}
\multicolumn{1}{|c|}{codec} & metric
& Slow-only& SlowFast & C2D & I3D & Swin & TPN\\
& & \cite{feichtenhofer2019slowfast} & \cite{feichtenhofer2019slowfast} & \cite{c2d} & \cite{i3d} & \cite{swin} & \cite{tpn}\\

\hline
H.264 & top-1 acc& -17.6& -12.3& -18.1 & -19.0 & -16.4& -18.8\\
     & top-5 acc& -17.1& -12.6& -18.5 & -18.9 & -16.8& -19.5\\
     & mean acc& -17.2& -13.0& -18.8 & -19.2 & -16.6& -19.6\\
    
    \cline{2-8}
    &VMAF &\multicolumn{6}{c|}{-5.6}\\

\hline
H.265 & top-1 acc& -16.2& -11.7& -16.9 & -16.6 & -15.4 & -18.1\\
     & top-5 acc& -16.2& -11.5& -16.5 & -16.9 & -15.8 & -18.3\\
     & mean acc& -16.0& -11.8& -16.8 & -17.1 & -15.6 & -18.4\\

     \cline{2-8}
    &VMAF &\multicolumn{6}{c|}{-5.1}\\
    
\hline

\end{tabular}}
\end{center}
\vspace{-10pt}

\end{table}

\begin{table}[tp]
\begin{center}
\caption{\label{}%
BD-Rate(\%) performance for object tracking}
{
\begin{tabular}{|c|c|c|c|c|c|}
\cline{1-6}
\multicolumn{2}{|c|}{~}& \multicolumn{4}{c|}{backbones}\\
\cline{1-6}
codec & metric &
KYS\cite{kys} & DiMP\cite{dimp} & ATOM\cite{atom} &PrDiMP\cite{prdimp}\\
\hline
H.264 & AUC & -15.5& -15.2 & -17.2 & -12.4\\

\cline{2-6}
    &VMAF &\multicolumn{4}{c|}{-6.3}\\

\hline

H.265 & AUC & -13.6& -12.4 & -16.6 & -12.0\\

\cline{2-6}
    &VMAF &\multicolumn{4}{c|}{-5.7}\\

\hline
\end{tabular}}
\end{center}
\vspace{-10pt}
\end{table}

We have conducted a comparative analysis of our preprocessing-enhanced video compression method against the conventional codec H.264 and H.265. Additionally, we apply the BD-Rate metric to quantify the percentage of bitrate savings while maintaining the same level of accuracy. Our default backbone networks for video action recognition and video object tracking are listed in Table 1 and Table 2, and we have trained their respective prerpocessing modules accordingly.

Evidently, our preprocessing-enhanced video compression method demonstrates a substantial enhancement in the trade-off between compression rate and accuracy when compared to the baseline approaches in the downstream action recognition task.
As depicted in Table 1, for the video action recognition task, we achieve significant improvements in BD-Rate performance across all four backbone networks. Notably, on H.264, we attain a maximum improvement of up to -17.6\%, and on H.265, we achieve a maximum improvement of up to -16.2\%. We have also conducted a comparative analysis of our method against traditional codecs in the context of the video object tracking task. Table 2 presents the rate-accuracy curves derived from various compression methods using the GOT-10k dataset. It is worth highlighting that our approach delivers superior rate-accuracy performance and achieves a maximum bitrate saving of up to 17.5\% when compared to the traditional codec baselines. The corresponding rate-accuracy curves are illustrated in Figure 4.

It is also worth noting that after applying our proposed preprocessor could let the output maintain a high degree of similarity to the original input. Furthermore, even after compression, we can achieve gains in perceptual-oriented metrics such as VMAF\cite{li2018vmaf}, which is human vision system-oriented. This is evident in both Table 1 and Table 2.

\begin{figure}[t]
\begin{center}
\begin{tabular}{cc}
\includegraphics[width=2.9in]{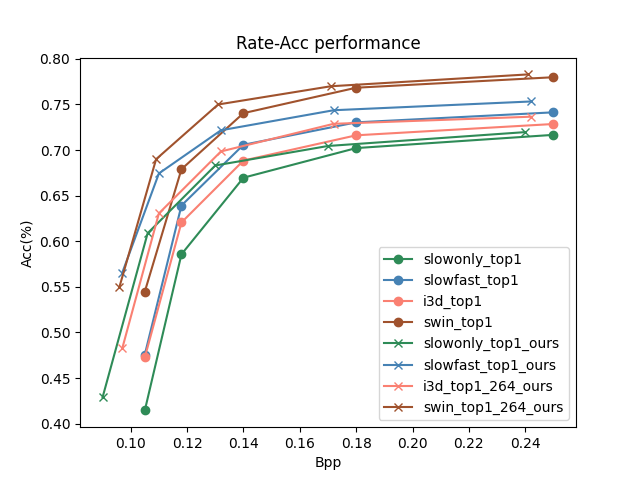} &
\includegraphics[width=2.9in]{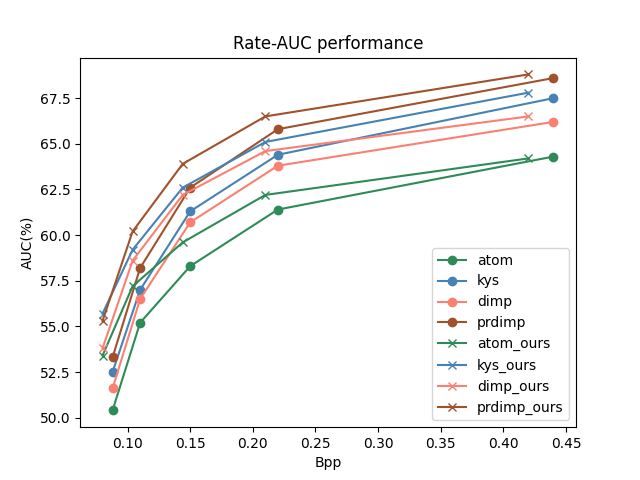} \\
{\small (a)} & {\small (b)}
\end{tabular}
\end{center}
\caption{\label{fig:rd}%
Rate-accuracy illustration plots for test results of video action recognition(a) and video object tracking(b).}
\end{figure}

\SubSection{Further Analysis}
To validate the effectiveness of the proposed preprocessing module, we conducted additional tests. In these tests, we directly employed compressed data from the video codec for fine-tuning the analyzer in downstream tasks. We utilized the same training strategy and anchor points as in the proposed approach. The trained analyzer was then compared to our proposed solution.
The experimental results, as shown in Table 3, indicate that fine-tuning with the compressed data does lead to some improvement in the analyzer's performance on compressed data. However, when compared to our proposed solution, it still lags significantly behind.

\begin{figure}[t]
\begin{center}
\includegraphics[width=6in]{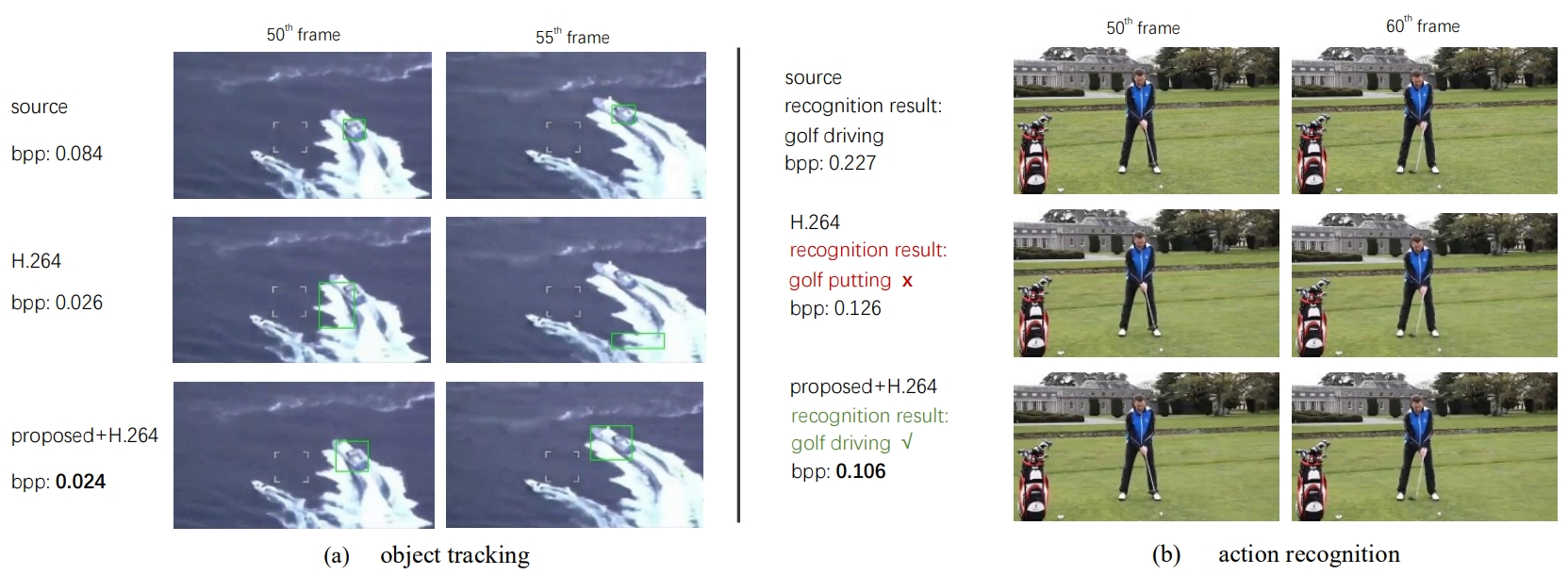} \\
\end{center}
\caption{\label{fig:vis}%
Visualization examples for proposed method outputs. (a) illustrates sample results for video object tracking, featuring frames from the Test000004 in the GOT-10k dataset. (b) presents sample results for video action recognition, showcasing images from the testing sequence 0u4c8Cel91U in the Kinetics400 dataset.}
\end{figure}

As shown in Figure 5, we provide some visualization results. As demonstrated in (a) for object tracking, it is evident that our proposed approach can successfully track the target even with a significant reduction in bitrate, whereas the comparative anchor method loses track of the target. In (b) for action recognition results, our proposed approach manages to have correct recognition result even with a substantial decrease in bitrate. In contrast, the anchor approach experiences a fatal in recognition with reduced bitrate. 


\begin{table}[tp]
\begin{center}
\caption{\label{}%
BD-Rate(\%) performance comparison for the fine-tuned version}
{
\begin{tabular}{|c|c|c|c|c|}
\cline{1-5}
& \multicolumn{4}{c|}{backbone}\\
\cline{2-5}
& \multicolumn{2}{c|}{Slow-only\cite{feichtenhofer2019slowfast}} & \multicolumn{2}{c|}{SlowFast\cite{feichtenhofer2019slowfast}}\\
\cline{1-5}
metric & fine-tuned &
ours & fine-tuned & ours\\
\hline
top-1 acc & -5.1 & -17.6 & -5.5 & -12.3\\
\hline
top-5 acc & -5.6 & -17.1 & -5.8& -12.6\\

\hline
\end{tabular}}
\end{center}
\vspace{-15pt}
\end{table}

\Section{Conclusion}
In the context of video machine vision, this paper introduces a video preprocessing framework aimed at enhancing the performance of downstream vision tasks under compression. We propose a neural network-based preprocessor to manipulate input videos, enabling the processed videos to better accommodate both video compression and machine vision tasks. Additionally, we introduce a virtual video codec to facilitate end-to-end training optimization. Experimental results demonstrate that our proposed approach yields significant bitrate savings compared to utilizing only standard codecs.

\vspace{300pt}

\Section{References}
\bibliographystyle{IEEEbib}
\bibliography{refs}

\end{document}